\documentclass[aps,prl,floatfix,twocolumn,tightenlines,10pt,amsfonts]{revtex4-1}
\usepackage[T1]{fontenc}
\usepackage{amsmath,amssymb}
\usepackage{graphicx}
\usepackage{amssymb}
\usepackage{color}
\usepackage{psfrag}
\usepackage{ifsym}
\usepackage{float}
\usepackage{soul} % \st = strikethrough
\usepackage[usenames,dvipsnames]{xcolor}
\usepackage[breaklinks=true]{hyperref}
\hypersetup{colorlinks = true, urlcolor = blue, linkcolor = blue, citecolor =  red}

\newcommand{\bra}[1] {\langle #1 |}
\newcommand{\ket}[1] {| #1 \rangle}

\newcommand{\ketbra}[1]{ | #1 \rangle\!\langle #1 |}

\begin{document}
\title{Quantum Hypercube States}
\author{L.~A. Howard$^{1}$, T.~J. Weinhold$^{1}$, F. Shahandeh$^{2}$, J. Combes$^{1}$, M.~R. Vanner$^{3}$, A.~G. White$^{1}$, and M. Ringbauer$^{4}$}
\affiliation{$^{1}$Centre for Engineered Quantum Systems, School of Mathematics and Physics, University of Queensland, Brisbane, Australia.\\
$^{2}$Department of Physics, Swansea University, United Kingdom.\\
$^{3}$QOLS, Blackett Laboratory, Imperial College London, London, United Kingdom.\\
$^{4}$Institut f{\"u}r Experimentalphysik, Universit{\"a}t Innsbruck, 6020 Innsbruck, Austria.}
           
\begin{abstract}   
\noindent
We introduce quantum hypercube states, a class of continuous-variable quantum states that are generated as orthographic projections of hypercubes onto the quadrature phase-space of a bosonic mode. In addition to their interesting geometry, hypercube states display phase-space features much smaller than Planck's constant, and a large volume of Wigner-negativity. We theoretically show that these features make hypercube states sensitive to displacements at extremely small scales in a way that is surprisingly robust to initial thermal occupation and to small separation of the superposed state-components. In a high-temperature proof-of-principle optomechanics experiment we observe, and match to theory, the signature outer-edge vertex structure of hypercube states.
\end{abstract}

\maketitle  

Non-Gaussian quantum states are commonly considered as a valuable resource for quantum-information processing~\cite{Mirrahimi2014}, tests of fundamental of physics~\cite{Bose1999,Marshall2003}, and sensing and metrology applications~\cite{Walther2004,Mitchell2004}. A crucial indicator for how useful a quantum state will be for these applications is how distinguishable it becomes from the initial state after a small displacement. This is closely related to the size of the smallest features in the state's phase space representation~\cite{Zurek2001}. Roughly speaking, two quantum states with smallest features occupying an area on the order of $d$ can become maximally distinguishable for displacements on the order of $\sqrt{d}$. Similarly, the rate of change in distinguishability in response to a displacement---a measure of the state's sensitivity to displacements---is a function of the size of the state's phase-space features. Quantum mechanics generally limits the size of these features to be at least on the order of $\hbar$, which is known as the shot-noise or standard quantum limit depending on the area of physics in which it arises.

\begin{figure}[b]
\includegraphics[width=0.55\columnwidth]{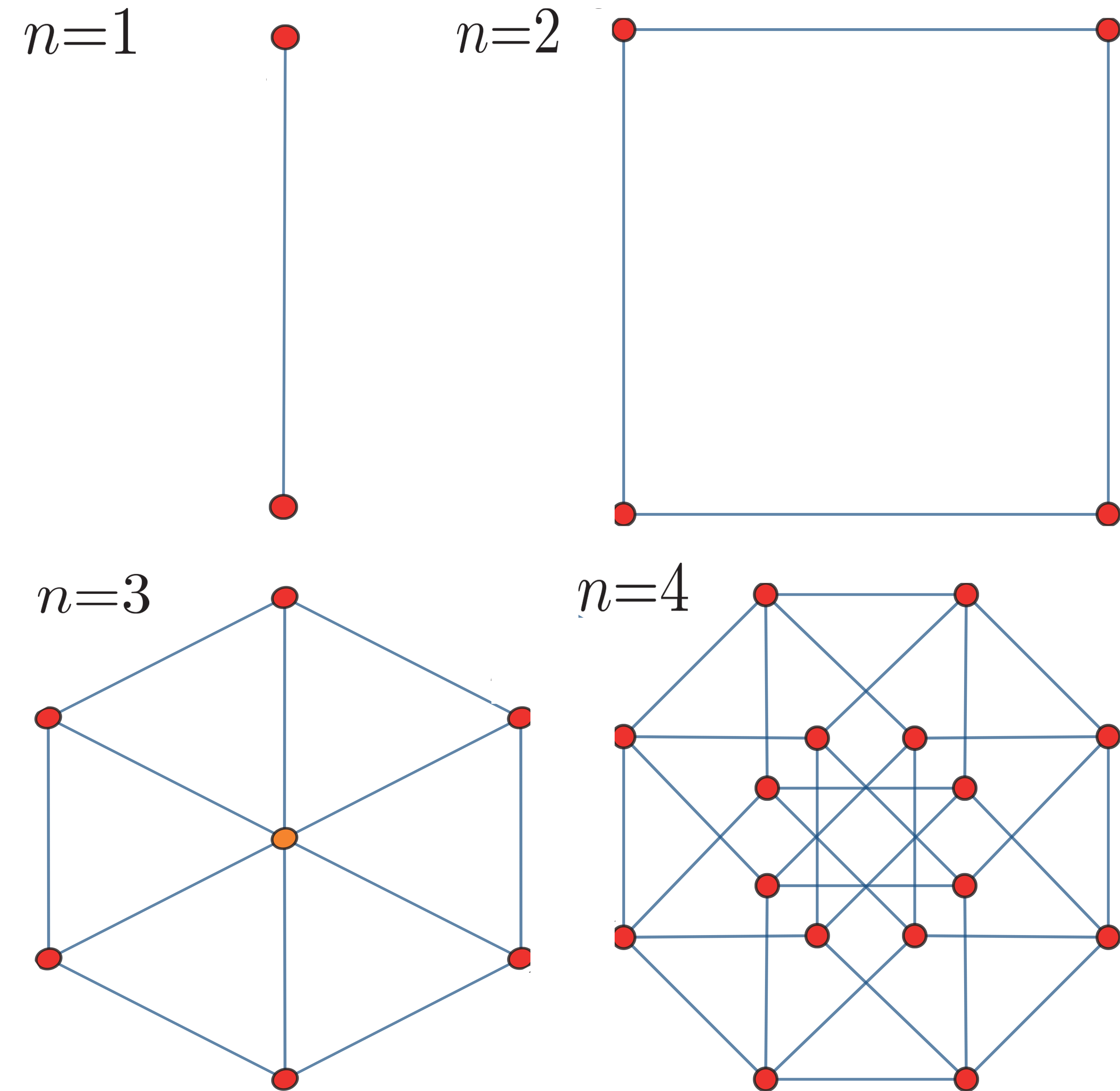}
\caption{Petrie polygon orthographic projections of first four \emph{n}-cubes. Red dots represents the projection of a single vertex onto a unique point in phase space, and orange dots the projection of two vertices onto the same point. \emph{n}=1 is the projection of a line-segment; \emph{n}=2 of a square; \emph{n}=3 of a cube; and \emph{n}=4 of a tesseract.}\vspace{-5mm}
\label{fig:Petrie}
\end{figure}

Yet, quantum theory also provides a way around this limit, and states such as the Schr\"odinger-cat state \cite{Leonhardt1997}---and the more-recently introduced compass state \cite{Zurek2001,Toscano2006}---show features at a scale below $\hbar$. States with such \emph{sub-Planck} features have thus attracted significant theoretical attention~\cite{Agarwal2004,Toscano2006,Dalvit2006,Ghosh2006,Roy2009,Choudhury2011,Praxmeyer2016} for their potential in sensing applications. However, most of the theory to date has focused on pure states, leaving open the question of how sensitive sub-Planck states will be under realistic conditions.
\begin{figure}[!b]
\includegraphics[width=\columnwidth]{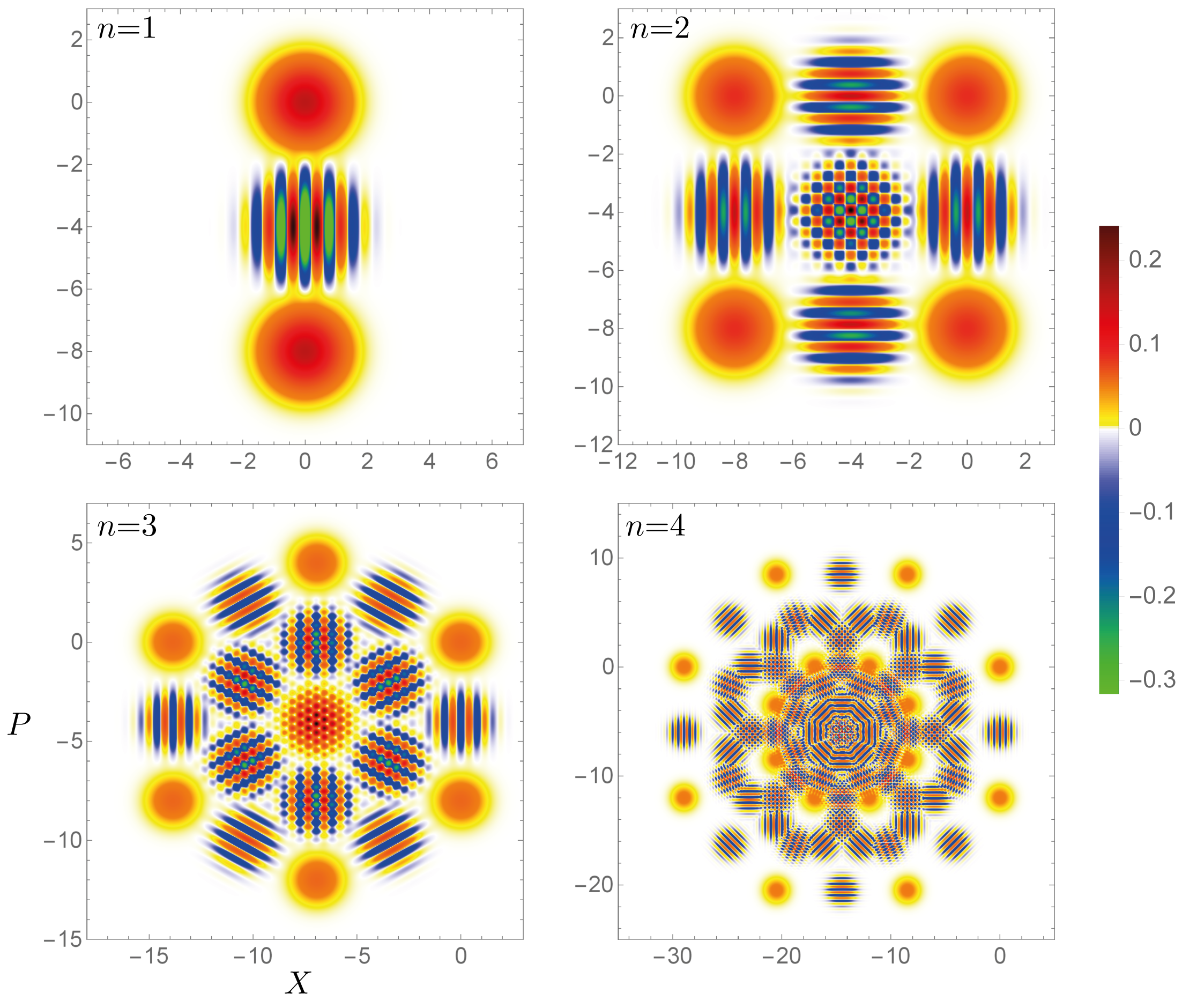}
\vspace{-7mm}
\caption{Density plots of the Wigner functions of the first four hypercube states. All axes are in terms of the ground state width.}\vspace{-5mm}
\label{fig:HypercubeStates}
\end{figure}

Here we introduce and study in detail a new class of non-classical states that we call \emph{hypercube states}. These states have an intriguing link to geometry in that they are obtained as Petrie-polygon orthographic projections of $n$-cubes~\cite{Davis2008}, see Fig.~\ref{fig:Petrie}, into phase space, where the polygon vertices correspond to the location of superposed coherent states, and interference fringes in the Wigner function are observed between every pair of vertices, see Fig.~\ref{fig:HypercubeStates}. This class of states in particular includes the Schr\"odinger-cat state and the compass state as the lowest-order special cases. All hypercube states exhibit sub-Planck phase-space features that decrease in size with the order of the state, making them an attractive candidate for quantum metrology. We show that hypercube states indeed become distinguishable under progressively smaller displacements with increasing order and, importantly, that this distinguishability is robust to variations in thermal occupation and/or interaction strength of the state for a wide range of realistic experimental parameters. Finally, building on the technique introduced in Ref.~\cite{Ringbauer2018} we discuss how quantum hypercube states can be prepared in a wide range of state-of-the-art optomechanical experiments. In a proof-of-principle demonstration we experimentally observed the signature of the second, third and fourth order hypercube states in the high-temperature regime.

\vspace{-0.2cm}
\section{Hypercube states}
\vspace{-0.2cm}
Mechanical hypercube states are engineered by the sequential application of an 
instrument or Kraus operator $\Upsilon$, defined as a superposition of the identity and a displacement, with intermittent phase-space rotations $R$ on the initial mechanical state $\rho_{\rm i}$.
The overall Kraus operator is thus given by
\vspace{-0.1cm}
\begin{equation}
\mathcal{Y}_{n} \propto \prod_{n} \Big(R\left(\pi/n\right) \Upsilon\Big).
\label{eq:general}
\end{equation}
Here, the rotations are due to the free evolution of the mechanics for durations of $t{=}T/2n$ that result in phase changes of $\theta {=} 2\pi t{/}T{=}\pi{/}n$, with $T$ being the period of the mechanical resonator and $n$ being the order of the desired hypercube state. Moreover, $\Upsilon{=}e^{-|\alpha|^{2}} \alpha \Big( \mathcal{D}(i\mu / \sqrt{2} ){-}1 \Big)/\sqrt{2}$ where $\lambda$ is the wavelength and $\alpha{\ll}1$ is the amplitude of the mediating coherent light field, $\mathcal{D}(\beta)$ ($\beta{\in}\mathbb{C}$) is the mechanical displacement operator and $\mu {=} 4 \pi x_{0} {/} \lambda$ is the interaction strength (equivalent to the optomechanical coupling strength in an optomechanical setting) in which $x_{0} {=} \sqrt{\hbar / m \omega}$ is the mechanical zero-point fluctuation. We note that $\Upsilon$ (and hence $\mathcal{Y}_{n}$) is obtained by conditioning on the outcome of a photon-counting measurement of the optical field after the radiation pressure interaction, and is thus non-unitary as detailed in the Supplemental Material (SM)~\cite{SM}. The Wigner function phase-space representation~\cite{Leonhardt1997} of the four lowest order hypercube states when the mechanics starts off from the vacuum are shown in Fig.~\ref{fig:HypercubeStates}.

\vspace{-0.3cm}
\section{$\ell_1$-norm sensitivity Analysis}
\vspace{-0.2cm}

\begin{figure*}[t]
\begin{center}
\includegraphics[width=\textwidth]{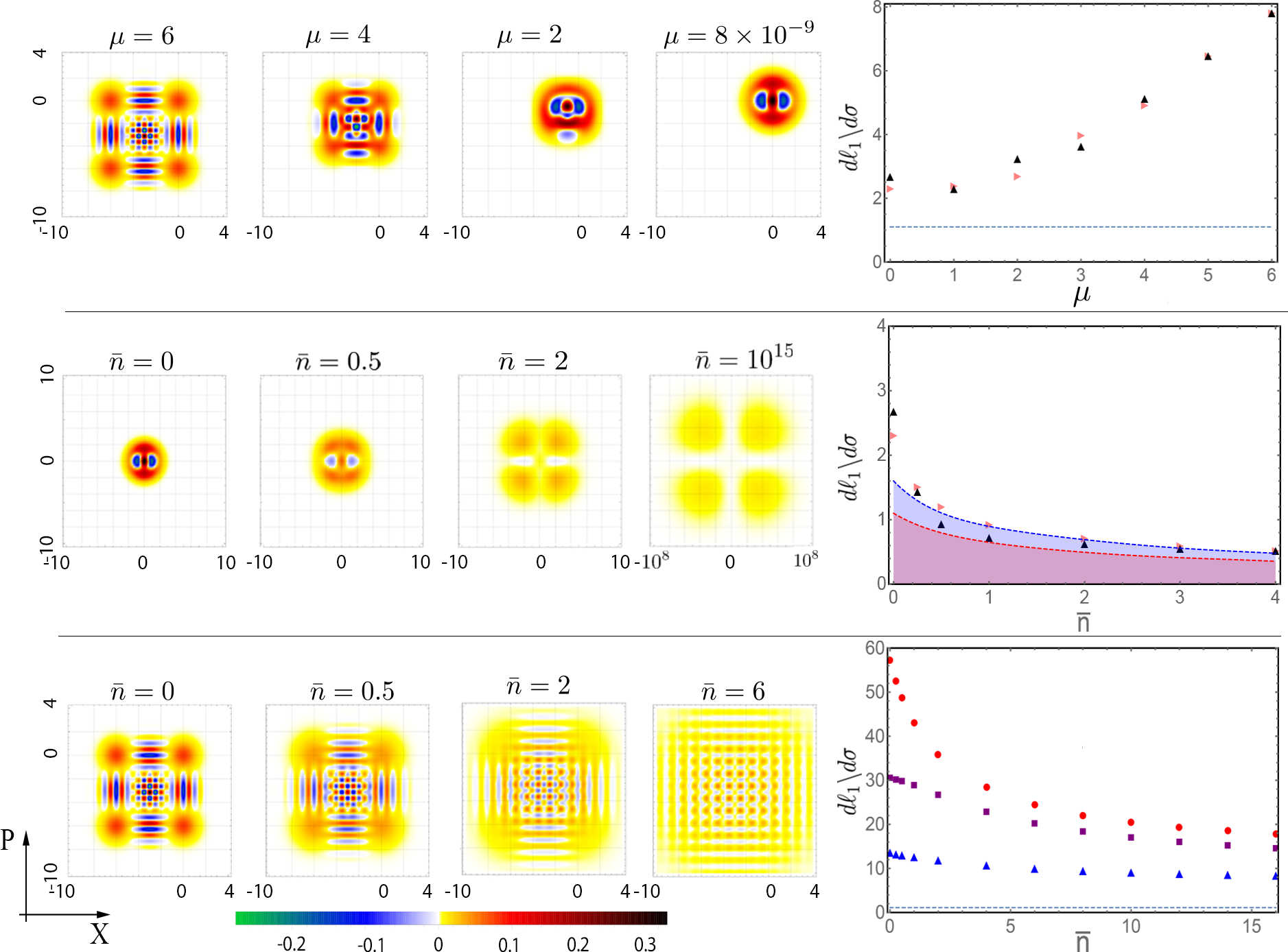}
\end{center}
\vspace{-5mm}
\caption{$n{=}2$ hypercube state for a mechanical resonator as produced by our scheme. \emph{Left-hand}: Theoretical density plots of the Wigner function in position-momentum ($X$-$P$) phase space. The colour bar on the bottom shows the value of the Wigner function where blue and green indicate areas of negative quasiprobability. All axes are in terms of the ground state width. \emph{Right-hand}: Corresponding $\ell_1$-sensitivity, $\mathrm{d}\ell_1/\mathrm{d}\sigma$, vs position/momentum displacement plots  (position, pink horizontal triangles) / (momentum, black vertical triangles). The blue dotted line indicates $\ell_1$-sensitivity of a coherent state at zero temperature.\textbf{Top row}. Cold temperature: small $\rightarrow$ large coupling ($\bar{n}{=}0$; $\mu{=}6 \rightarrow \mu{=} 8 {\times} 10^{-9}$). 
\textbf{Middle row} small coupling: cold $\rightarrow$ hot temperature ($\mu{=} 8 {\times} 10^{-9}$; $\bar{n}{=}0 \rightarrow \bar{n}{=} 10^{15}$; the colours in the last panel are scaled by $10^{8}$ in accordance with the axis scaling).The $\ell_1$-sensitivity plot shows that even hypercube states with vanishing separation of coherent states are more sensitive than thermal states (red shaded region) and at least as sensitive as squeezed (3dB) thermal states (blue shaded region) with the same mean phonon number and more sensitive than a ground state coherent state for $\bar{n}\lesssim 0.5$. \textbf{Bottom row} Large coupling: cold $\rightarrow$ warm temperature ($\mu{=} 6$; $\bar{n}{=}0 \rightarrow \bar{n}{=} 6$). $n{=}2$ hypercube states are shown by blue triangles; for comparison, $n{=}3$ \& $n{=}4$ hypercube states are shown respectively by purple squares and red circles. Note in the right plot, $\mu{=} 12$, was used to ensure clear separation of coherent states across all the orders of hypercube state.}\vspace{-5mm}
\label{fig:prog}
\end{figure*} 

In order to quantify how quickly a displaced quantum state $\rho_{\sigma}$ becomes distinguishable from the initial state $\rho$ for small displacements $\sigma$, we use the $\ell_1$-distance between the corresponding Wigner functions $W_{\rho}$ and $W_{\rho_\sigma}$,
\begin{equation}
\ell_1{:=}\| W_{\rho_{\sigma}}{-}W_{\rho} \|_{1}{=}\int_{-\infty}^{\infty}\mathrm{d}x \mathrm{d}p \; |W_{\rho_{\sigma}}(x,p){-}W_{\rho}(x,p)|, \nonumber
\end{equation}
and consider the rate of change $\mathrm{d}\ell_1/\mathrm{d}\sigma$ of this distance at the point $\sigma{=}0$. Importantly, this measure, denoted as $\ell_1$-{\it sensitivity}, can easily be applied in the \emph{thermal regime} which is a crucial element of this work.

Different regimes of operation can be parameterized by the interaction strength $\mu$ and the temperature of the mechanical mode as measured by the average phonon number $\bar{n}$.
We are particularly interested in the $\ell_1$-sensitivity of hypercube states in three limiting cases:  (i) \emph{cold, large-coupling}: mechanical resonator in the ground state, $\bar{n} {=} 0$, and $\mu {=}$6, 8, and 12, for the second-, third-, and fourth-order hypercube states, respectively; (ii) \emph{cold, small-coupling}: mechanical resonator in the ground state, $\bar{n} {=} 0$, and $\mu {=} 8 {\times} 10^{-9}$; and lastly (iii) \emph{hot, small-coupling}: mechanical resonator in a thermal state with $\bar{n} {\sim} 10^{15}$, and small interaction, $\mu {=} 8 {\times} 10^{-9}$, motivated by the proof-of-principle experiment presented at the end of the manuscript.
The fourth limiting regime of hot-large coupling is less experimentally relevant, because experiments that achieve large coupling between the optical field and resonator can typically cool the resonator very efficiently using laser cooling methods. Nevertheless we do consider how the $\ell_1$-sensitivity changes as a strongly coupled resonator warms up, effectively moving from the cold-large towards the hot-large coupled regime, see bottom row of Fig.~\ref{fig:prog}.

Our theoretical  analysis of the $\ell_1$-sensitivity of a second-order hypercube state transitioning between the above described regimes is summarised in Fig.~\ref{fig:prog}.
In the top row of Fig.~\ref{fig:prog} we show how increasing the interaction strength $\mu$ to transition from the small- to the large-coupling regime at low temperature leads to smaller phase-space features and an $\ell_1$-sensitivity that increases quadratically with $\mu$ for $\mu \lesssim 3$ and linearly for larger $\mu$, see SM for details~\cite{SM}. Note, however, that for any value of $\mu$ hypercube states are more sensitive to displacements than a coherent state and maintain significant Wigner negativity (see Fig.~S4). Interestingly, for very small interaction strengths the symmetry around the position axis breaks, meaning that at some values the state is more sensitive to displacements in momentum than in position, or vice versa; see SM for details~\cite{SM}.

The middle row of Fig.~\ref{fig:prog} shows the transition between the small-coupling regimes and that as the temperature increases Wigner negativity and smaller scale features disappear and the outer-edge vertex structure becomes the dominant feature. As a result, the $\ell_1$-sensitivity of the states decreases exponentially with temperature, see SM for details~\cite{SM}. Interestingly, however, hypercube states show an $\ell_1$-sensitivity at least as high as 3dB-squeezed coherent states of the same average phonon number even in the regime where $\mu\ll 1$. For low temperatures of $\bar{n}\lesssim 0.25$, hypercube states are more sensitive than the corresponding squeezed states.

In the bottom row of Fig.~\ref{fig:prog} we increase temperature in the cold large-coupling regime. Here we see that periodic and symmetric sub-Planck features are visible for $0 {<} \bar{n} {\leqslant} 6$. In fact, they exist even at much higher temperatures. 
The $\ell_1$-sensitivity plot on the right side of the row highlights both that, in this regime, the $\ell_1$-sensitivity is robust to temperature increases, and that it increases substantially with the order of the hypercube. These results apply equally to displacements along the position or momentum axis.

\vspace{-0.5cm}
\section{Experimental Method and Results}
\vspace{-0.4cm}
Our experimental method utilises a series of $n$ interactions between a single photon and mechanical resonator~\cite{Ringbauer2018}, followed by single photon detection to create an $n^\text{th}$ order hypercube state within the mechanical resonator. Prior to the interaction, each single photon is in a quantum superposition of being incident, and not incident on a mechanical resonator. Interaction of the mechanical resonator and a photon thereby puts the mechanical resonator in a quantum superposition of receiving, and not receiving a momentum kick due to the radiation pressure applied by the photon incident on the resonator. This interaction effectively applies $\Upsilon$ and hence a series of photons prepared this way, with time $t{=}T/2n$ between successive photons acts to apply Eq.~\eqref{eq:general} to the state of mechanical resonator. A schematic of the setup, with further details in the caption is shown in Fig.~\ref{fig:schematic}.

\begin{figure}[!t]
\begin{center}
\includegraphics[width=0.8\columnwidth]{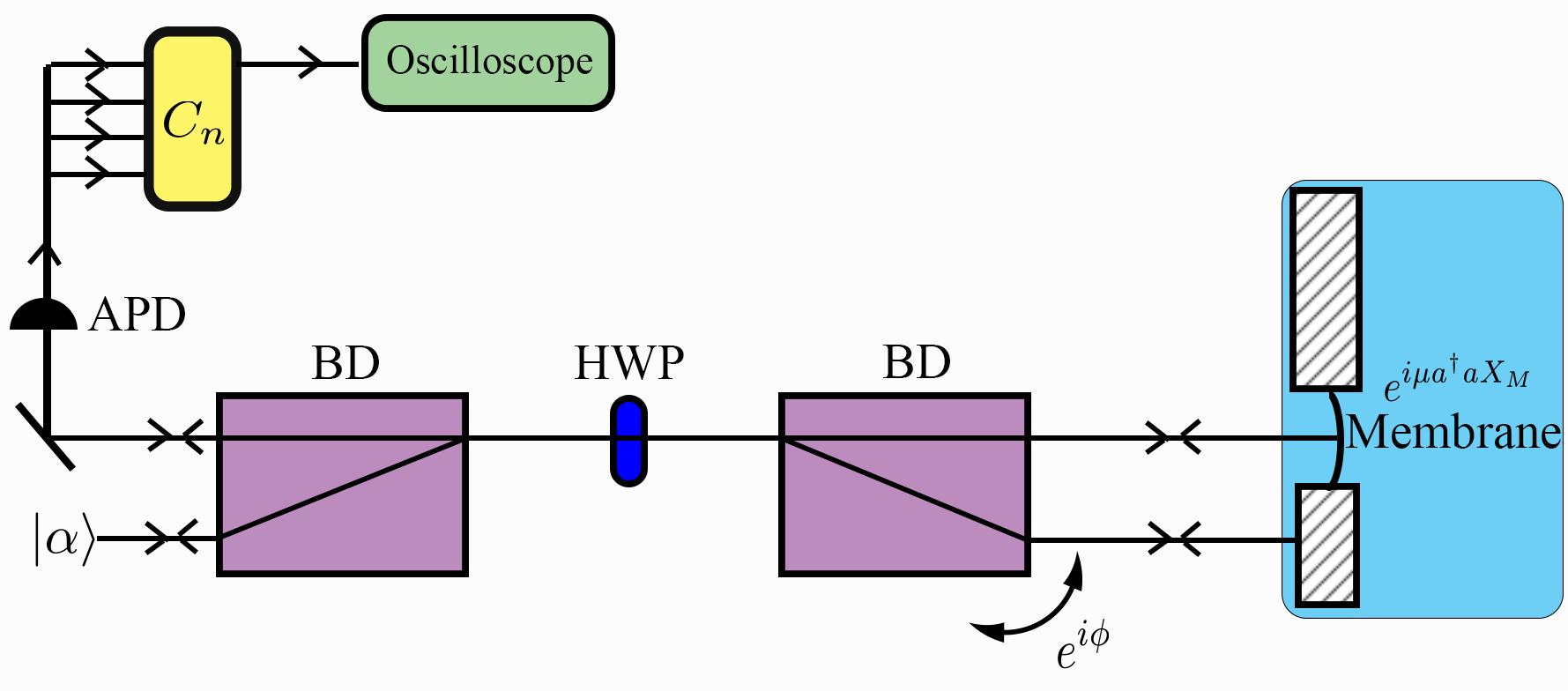}
\end{center}
\vspace{-2em}
\caption{A weak coherent state $|\alpha \rangle$ is inserted and split using a half wave plate (HWP) and a beam displacer (BD) into two arms of a folded interferometric setup. The optical field in the top arm interacts with a membrane mechanical resonator via radiation pressure, described by $e^{i\mu a_{1}^{\dagger}a_{1}X_{M}}$. Here $a$ and $a^{\dagger}$ are the annihilation and creation operators, $\mu$ is the optomechanical coupling strength and $X_{M}$ is the mechanical position operator. The bottom arm interacts with the static frame of the membrane, obtaining a controllable phase shift $e^{i \phi}$, which we set to $\phi {=}\pi$. Post these interactions, the optical fields in both arms are interfered, split into the polarisation components and detected in an avalanche photon detector (APD). The detector signal is then split into two, three or four paths depending on whether detection of a second, third or fourth order hypercube is being made; with each path past the first adding a time delay of $T/2n$. Coincidence counting between the paths is made at $C_{n}$, and upon detection of a coincidence event an oscilloscope is triggered to record a trace of the membrane's position from the back (not shown).}\vspace{-4mm}
\label{fig:schematic}
\end{figure}

We now introduce the experimental results from implementing our method in the hot small-coupling regime with a 100 ng (${\simeq} 10^{16}$ atoms) mechanical resonator. To ensure a large physical displacement that can be easily measured, the resonator is driven with a piezo (in order to create a Gaussian state in phase space the piezo drive voltage is Chi-distributed due to the drive voltage coupling to $\sqrt{X^{2} {+} P^{2}}$). This gives the resonator an effective thermal occupation of $\bar{n} {=} 10^{15}$. The mechanical resonator was coupled to 795 nm light, with the application of the measurement operator $\mathcal{Y}_{n}$ heralded by $n$ sequential single-photon detections. Detection of photons separated by $T/2n$ was accomplished by splitting the electronic signal from the photon detector into multiple paths of varying lengths and then looking for coincidences between the paths, see SM for details~\cite{SM}.
\begin{figure}[!t]
\begin{center}
\includegraphics[width=0.95\columnwidth]{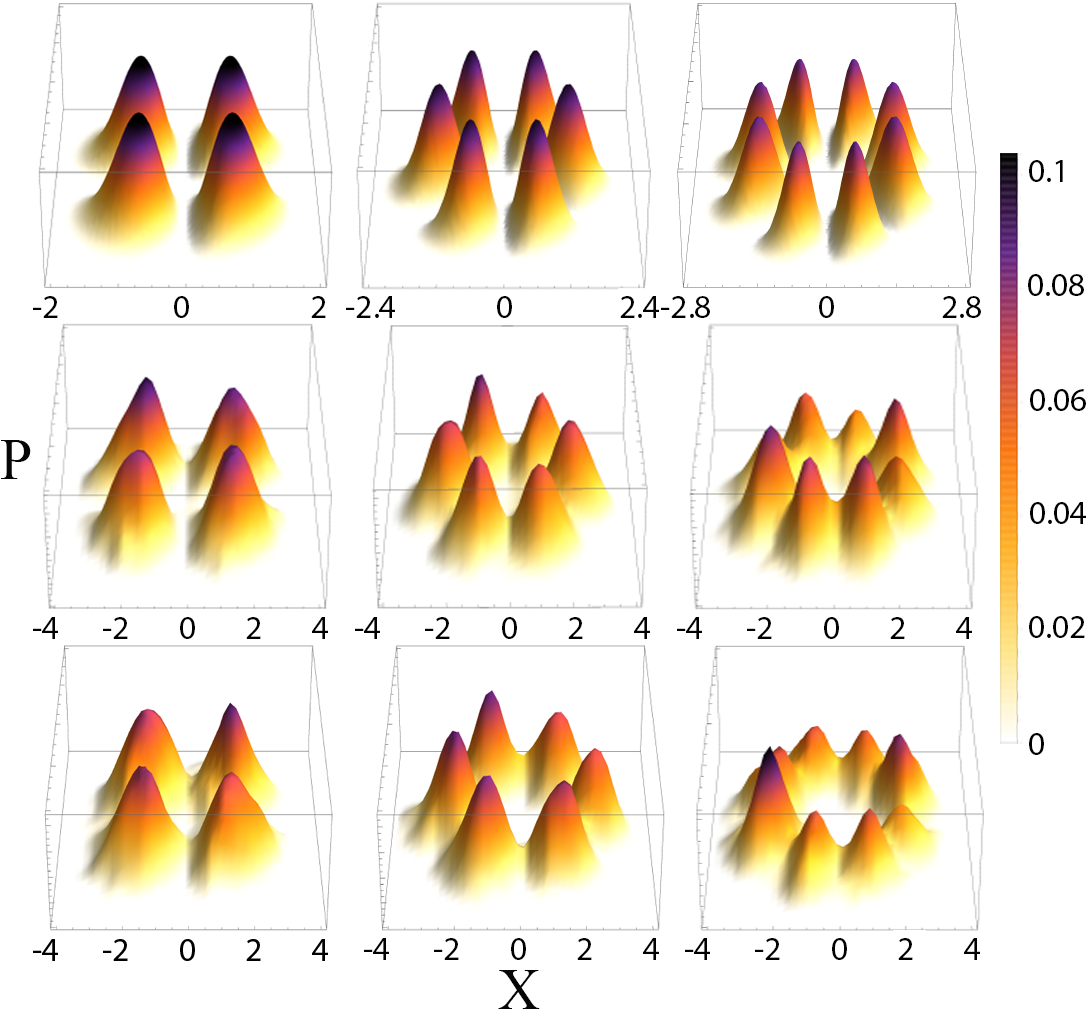}
\end{center}
\vspace{-2em}
\caption{Normalised probability densities of hypercube states; left to right columns correspond to second-, third- and fourth- order hypercube states. \textbf{Top row}. Ideal densities, from equation~\ref{eq:general}: populations match the outer-ring of vertices of the Petrie projections in Fig.~\ref{fig:Petrie}; quantum features such as sub-Planck structure are absent due to high temperature. \textbf{Middle row}. Densities predicted from a model which accounts for experimental drift, timing uncertainties and counting statistics, see SM~\cite{SM}. Note the predicted variations in peak height, width and shape. \textbf{Bottom row}. Measured probability densities, the Bhattacharyya coefficient (a classical analogue of the fidelity) between the middle row of density plots and the experimentally measured density plots are $0.95\pm0.02$, $0.90\pm0.02$ and $0.82\pm 0.02$ for the second, third and fourth-order hypercube states respectively. The uncertainties represent one standard deviation obtained from the statistical uncertainties in determining the values of X and P. The signature of hypercube states in the thermal regime---the ring of outer-edge-vertices---is clearly visible in all plots. All axes are in units of $\lambda_{R}/4=158.2nm$, where $\lambda_{R}$ is the readout wavelength used to measure the mechanical resonators phase space position.\vspace{-7mm}}
\label{fig:exp}
\end{figure}

Figure~\ref{fig:exp} displays, and compares to theory, our experimental measurements of hypercube states of orders $n{=}2$, $n{=}3$, and $n{=}4$ in the hot small-coupling regime. The top row of Fig.~\ref{fig:exp} shows the theoretical expectation in the ideal case of no experimental limitations: probability densities that reproduce the outer-ring of vertices of the Petrie projections in Fig.~\ref{fig:Petrie}, and that do not contain quantum features, or sub-Planck structure---as expected due to the high temperature. These plots are generated by applying $\mathcal{Y}_{n}$ of equation~\ref{eq:general}. The middle row of Fig.~\ref{fig:exp} adds modeling of the major experimental limitations present, namely: drift in the interferometer setting, $\phi$; experimental timing uncertainties when applying $R(\pi/2n)$; and variations from the mean due to counting statistics and displays notable variations in the height, width, and shape of the peaks in the probability densities; see SM for details~\cite{SM}. The bottom row shows our measured results which show the same features---there is excellent agreement between the model predictions and our experiments (see the caption of Fig.~\ref{fig:exp} quantification of agreement). 

Hypercube states establish a strong connection to geometry for continuous-variable quantum states, as well as providing an avenue for sub-Planck sensing over a surprisingly large range of experimental parameters. Given their potential, and the pivotal role geometry has played in the field of discrete-variable quantum information processing for problems such as measurement-based quantum computing and quantum error correction; we expect hypercube states will inspire diverse applications in fields such as quantum sensing, quantum information theory and quantum foundations~\cite{Anastopoulos2015, Marshall2003, Bassi2013}. To achieve these applications will require an amalgamation of current technologies for precision quadrature measurements at the sub-Planck scale~\cite{Vanner2015,Jacobs2017,Shahandeh2017}, and cooling a mechanical resonator to close to the ground state~\cite{Rossi2018}.\\

\noindent \textbf{Acknowledgements}. We acknowledge support from: the Australian Research Council (ARC) via the Centre of Excellence for Engineered Quantum Systems (EQUS, CE170100009), a Discovery Project for MRV (DP140101638) and a Discovery Early Career Researcher Award for JC (DE160100356); the Engineering and Physical Sciences Research Council (EP/N014995/1); and the University of Queensland by a Vice-Chancellor's Senior Research and Teaching Fellowship for A.G.W. M.R. acknowledges funding from the European Union's Horizon 2020 research and innovation programme under the Marie Sk\l{}odowska-Curie grant agreement No 801110 and the Austrian Federal Ministry of Education, Science and Research (BMBWF). F.S. acknowledges support and resources provided by the S\^{e}r SAM Project at Swansea University, an initiative funded through the Welsh Government's S\^{e}r Cymru II Program (European Regional Development Fund). We acknowledge the traditional owners of the land on which the University of Queensland is situated, the Turrbal and Jagera people.

\section*{Supplementary Material}
\renewcommand{\theequation}{S\arabic{equation}}
\renewcommand{\thesection}{S\arabic{section}}
\renewcommand{\thefigure}{S\arabic{figure}}
\setcounter{equation}{0}
\setcounter{section}{0}
\setcounter{figure}{0}

\noindent \textbf{Optomechanical interaction.}
Here we briefly review the optomechanical interaction and the derivation of the Kraus operators used in the main text adapted from Ref.~\cite{Ringbauer2018}. The radiation-pressure interaction between a mechanical resonator with position operator $\hat X_M$ and an optical field with annihilation operator $\hat a_1$ is described by the two-mode unitary operation $e^{i\mu \hat a_1^\dagger \hat a_1 \hat X_{M}}$. Here $\mu {=} 4 \pi x_{0} {/} \lambda$ is the interaction strength for a resonator with zero-point fluctuations of $x_{0} {=} \sqrt{\hbar / m \omega}$ and an optical wavelength $\lambda$.

\begin{figure}[h!]
\begin{center}
\includegraphics[width=0.8\columnwidth]{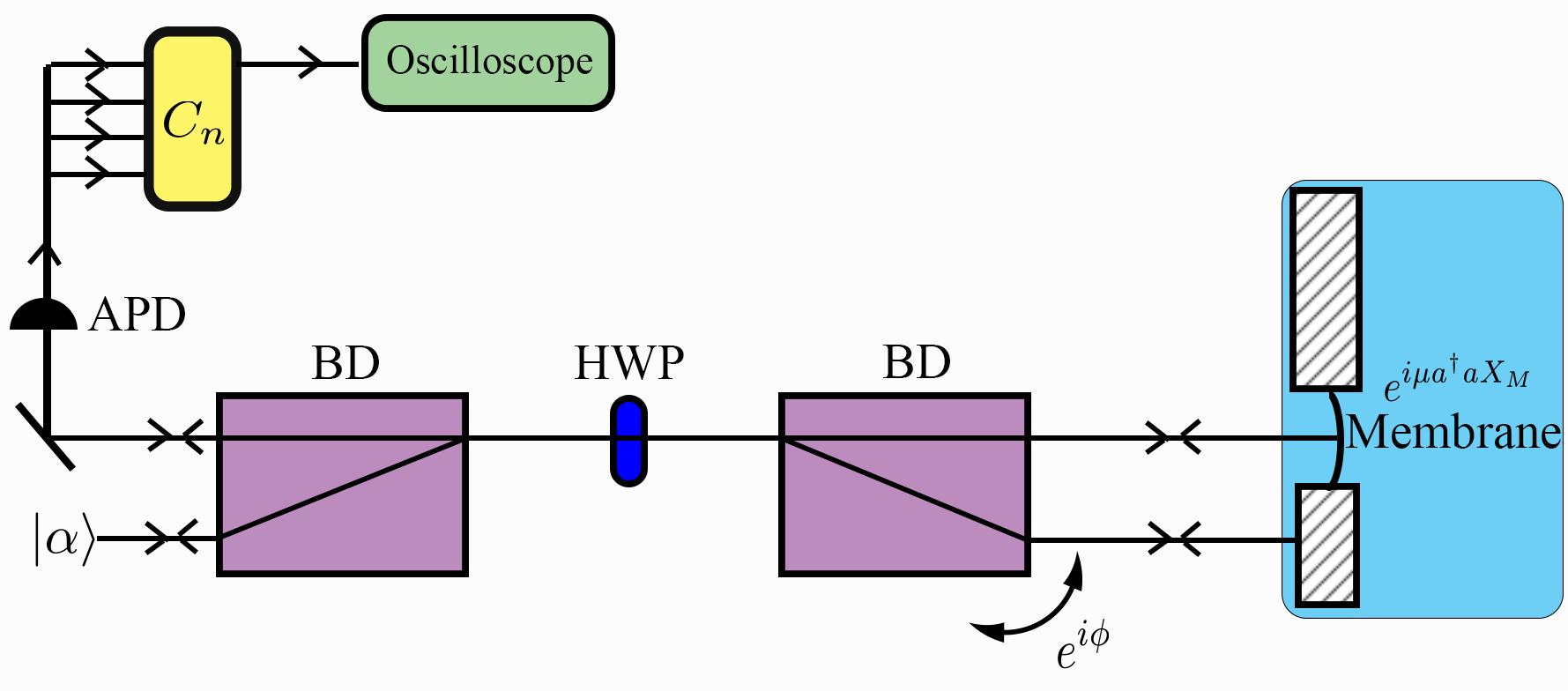}
\end{center}
\vspace{-2em}
\caption{(Fig.~4 of the main text) 
A weak coherent state $|\alpha \rangle$ is inserted and split using a half wave plate (HWP) and a beam displacer (BD) into two arms of a folded interferometric setup. The optical field in the top arm interacts with a membrane mechanical resonator via radiation pressure, described by $e^{i\mu a_{1}^{\dagger}a_{1}X_{M}}$. Here $a$ and $a^{\dagger}$ are the annihilation and creation operators, $\mu$ is the optomechanical coupling strength and $X_{M}$ is the mechanical position operator. The bottom arm interacts with the static frame of the membrane, obtaining a controllable phase shift $e^{i \phi}$, which we set to $\phi {=}\pi$. Post these interactions, the optical fields in both arms are interfered, split into the polarisation components and detected in an avalanche photon detector (APD). The detector signal is then split into two, three or four paths depending on whether detection of a second, third or fourth order hypercube is being made; with each path past the first adding a time delay of $T/2n$. Coincidence counting between the paths is made at $C_{n}$, and upon detection of a coincidence event an oscilloscope is triggered to record a trace of the membrane's position from the back (not shown).}\vspace{-4mm}
\label{fig:schematic}
\end{figure}

As shown in the Fig.~\ref{fig:schematic} above (Fig.~4 of the main text), the mechanical resonator is embedded in a balanced interferometer with a static phase-shift of $\phi=\pi$ in the other arm. An incoming coherent light field is split into two equal components of magnitude $\alpha$, of which one half interacts with the mechanical resonator. Subsequently, the two optical fields are interfered on a balanced beam splitter to close the interferometer, followed by photon counting on the output ports of the final beam splitter. Upon conditioning on detecting exactly one photon in the first output mode of the final beam splitter and none in the second, we obtain

\begin{equation}
\label{Eq:UpsilonBraUKet}
\begin{split}
\Upsilon \hat{\rho}_{\rm i}\Upsilon^\dag = {\rm Tr}_{12}\Big[\Big(\ketbra{0}{1} \otimes \ketbra{1}{2}\Big) B_{12} e^{i\mu a_1^\dagger a_1 X_M} \Big(\ketbra{\alpha}{1}\\
\otimes \ketbra{\alpha e^{i\phi}}{2}
\otimes \hat{\rho}_{\rm i}\Big) e^{-i\mu a_1^\dagger a_1 X_M}B_{12}^\dag\Big]\, .    
\end{split}
\end{equation}
Here $B_{12}$ is the beamsplitter operator and subscripts 1 and 2 refer to the optical modes of the interferometer.
Also, $\hat{\rho}_{\rm i}$ refers to the initial mechanical state.
Thus equivalently we can write~\cite{Ringbauer2018}
\begin{equation}
\label{Eq:UpsilonBraUKet2}
\Upsilon = {_{2}\bra{0}}{_{1}\bra{1}} B_{12} e^{i\mu a_1^\dagger a_1 X_M} \ket{\alpha}_{1} \ket{\alpha e^{i\phi}}_{2}\, .
\end{equation}
Expanding this equation we find
\begin{equation}
\Upsilon = e^{-|\alpha|^{2}} \alpha \left(e^{i\mu X_M} + e^{i\phi}\right)/\sqrt{2} \, ,
\end{equation}
so that by setting $\phi=\pi$ we find the expression for $\Upsilon$ as given in the main text.
We note that, since this is a conditional operation, the Kraus operator $\Upsilon$ corresponds to a non-unitary map.

\noindent \textbf{Theoretical modeling of Wigner functions}. To find the Wigner function after applying Eq.(2) to an initial state $\rho$, we write the final state using symmetric ordering and then make the substitutions $b {=} (X {+} iP)/\sqrt{2}$ and $b^{\dagger} {=} (X {-} iP)/\sqrt{2}$, where $X$ and $P$ are the position and momentum coordinates in the Wigner function. Before elaborating on this process we rewrite $\Upsilon {=} e^{{-}|\alpha|^{2}} \alpha (e^{i\mu X_{M}} + 1)/\sqrt{2}$ using the displacement operator $\mathcal{D}(\beta) {=} e^{\beta b^{\dagger} {-} \beta^{*}b}$ for a displacement of the mechanical mode with bosonic creation and annihilation operators $b^\dag$ and $b$ by $\beta\in\mathbb{C}$. To do this we note that $\phi {=} 0$, $X_{M} {=} (b^{\dagger} {+} b)/\sqrt{2}$, and the overall scalar factors can be ignored as the final state will be renormalised. We can then rewrite $\Upsilon$ as;

\begin{align}
\Upsilon=&e^{i\mu X_{M}}-1\\ \nonumber
=&e^{(i\mu/\sqrt{2})(b^{\dagger}+b)}-1\\ \nonumber
=&\mathcal{D}(i\mu /\sqrt{2})-1.
\end{align}
This shows that our measurement operator is equivalent to applying a superposition of the displacement operator and the identity and allows us to use known algebra regarding the displacement operator, e.g.\ $\mathcal{D}(\beta)|0\rangle {=} |\beta \rangle$ and $\mathcal{D}^{\dagger}(\beta) {=} \mathcal{D}({-}\beta)$.\\

Writing our initial state $\rho$ in symmetric ordering as $\left \{e^{-\lambda b^{\dagger}b} \right \}_{0}$, we then need to apply the appropriate number of $\mathcal{D}(i\mu /\sqrt{2}) {+} 1$ and $R(\theta)$ operations in Eq.~(2), whilst always ensuring that the state is written symmetrically. To calculate the second-order hypercube state we perform the following calculation,
\begin{align*}
&(\mathcal{D}(i\mu / \sqrt{2})-1)R(\frac{\pi}{2})(\mathcal{D}(i\mu / \sqrt{2})-1)\rho_{\rm i}\\
& (\mathcal{D}^{\dagger}(i\mu / \sqrt{2})-1)R^{\dagger}(\frac{\pi}{2})(\mathcal{D}^{\dagger}(i\mu / \sqrt{2})-1)=\\
&(\mathcal{D}(i\mu / \sqrt{2})-1)R(\frac{\pi}{2})(\mathcal{D}(i\mu / \sqrt{2})-1)\rho_{\rm i}\\
& (\mathcal{D}(-i\mu / \sqrt{2})-1)R(-\frac{\pi}{2})(\mathcal{D}(-i\mu / \sqrt{2})-1).
\end{align*}

\noindent This function has 16 terms. The third- and fourth- order compass states have 64 and 256 terms respectively. An exemplary calculation of the application of the displacement operator to the ground state once, i.e., $\mathcal{D}(\beta)\left \{e^{-\lambda b^{\dagger}b} \right \}_{0}$ using the general ordering theorem developed in Ref.~\cite{Shahandeh2012} is shown below:
\begin{align*}
\mathcal{D}(\beta) &\left  \{e^{-\lambda b^{\dagger}b} \right \}_{0} = e^{ (\beta b^{\dagger}- \beta^{*}b)} \left \{e^{-\lambda b^{\dagger}b} \right \}_{0}\\ 
&\quad = e^{-|\beta |^{2}/2}e^{ \beta b^{\dagger}}e^{-\beta^{*} b} \left \{ e^{-\lambda b^{\dagger}b}\right \}_{0}\\ 
&\quad = e^{-|\beta |^{2}/2}e^{ \beta b^{\dagger}}\left \{ e^{-\beta^{*} b} e^{-\lambda (b^{\dagger} {-} \beta^{*}/2)b}\right \}_{0}\\
&\quad = e^{-|\beta |^{2}/2}\left \{ e^{ \beta b^{\dagger}} e^{-\beta^{*} b} e^{-\lambda (b^{\dagger} {-} \beta^{*}/2)(b-\beta/2)}\right \}_{0}
\end{align*}
\begin{figure}[h!]
\begin{center}
\includegraphics[width=\columnwidth]{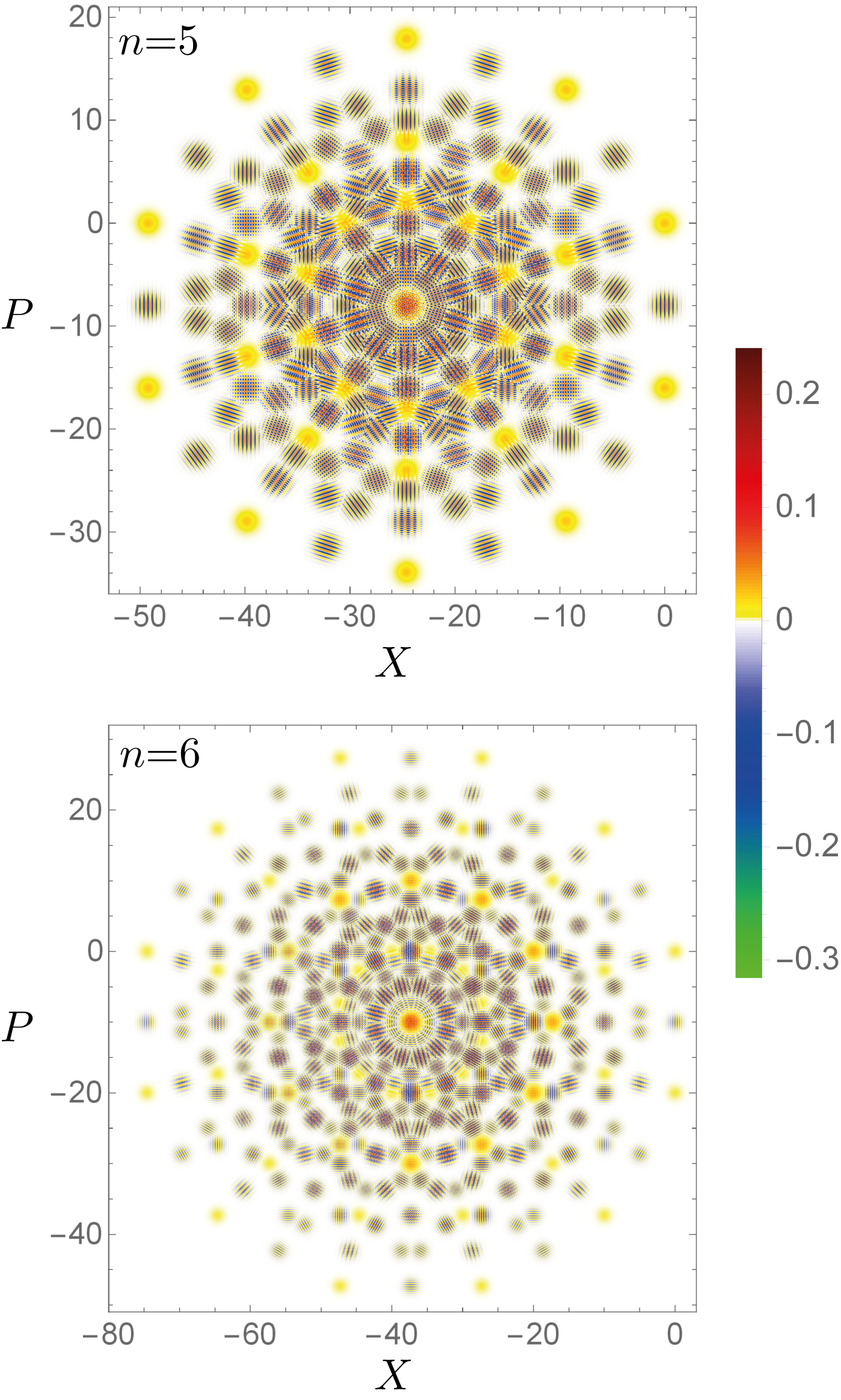}
\end{center}
\vspace{-6mm}
\caption{Density plots of the hypercube state Wigner functions for $n=5$ and $n=6$, generated from an initial mechanical ground state respectively for with $\mu=16$ and 20. Colour coding as for Fig.~2. \vspace{-5mm}}
\label{Fig:HypercubeStates2}
\end{figure}

\noindent \textbf{Modelling of experimental uncertainties}. A model was constructed to ascertain if our experimental results could be obtained from ideal hypercube states in our experimental regimes by modelling the experimental uncertainties present, namely, drift in the interferometer setting, $\phi$; experimental timing uncertainties when applying $R(\pi/2n)$; and variations from the mean due to counting statistics. The resultant states after accounting for these experimental uncertainties are shown in Fig~5 and show excellent agreement with our measured results. 

More detail on the modelling of each experimental uncertainty is now provided. Experimental drift of $\phi$ between measurements mean that each state preparation is averaged over different values of $\phi$. Having recorded the value of $\phi$ for each measurement, we can compute the range of $\Upsilon$, average them, and apply the average in Eq.~(2). This error was kept small by rejecting datasets with drifts larger than $10^{\circ}$.  Timing Uncertainty between photon detections of ${\pm} 15$ ns resulted in uncertainty in the argument of rotation operator $R(\pi/2n)$ in Eq.~(2). This timing uncertainty, $\delta t$, was approximated by averaging two states with $R(\pi/2n+\delta t\pi/T)$ and $R(\pi/2n {-} \delta t \pi/T)$. Finally, a normalised probability density plot from a number of samples (equal to that in the experiment) from this distribution, taking into account Poissonian counting statistics.\\

\begin{figure}[!t]
\begin{center}                
\includegraphics[width=\columnwidth]{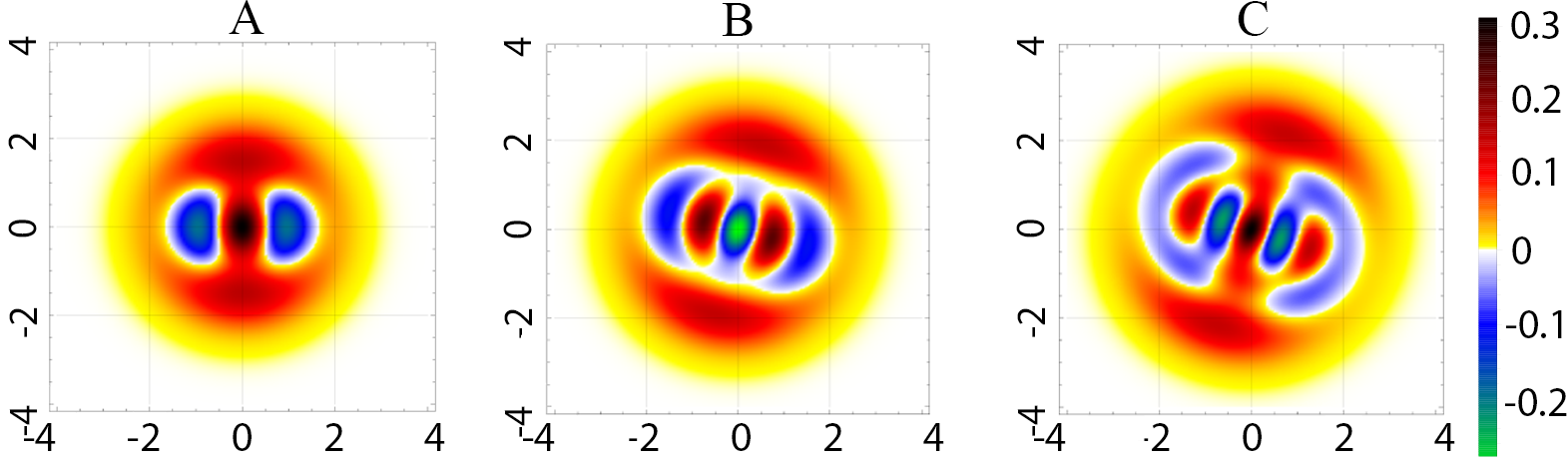}
\end{center}
\vspace{-5mm}
\caption{Theoretical density plots of the Wigner-function for A) second- B) third-, and C) fourth-order hypercube states for $\bar{n}{=}0$ and $\mu {=} 8 {\times} 10^{-9}$ in units of the ground state width. Each state exhibits a minimum Wigner negativity of ${\leq}{-}0.2$. The rotations observed in B) and C) are due to additional phases acquired from multiple applications of the displacement operator. \vspace{-5mm}}
\label{fig:lowtemplowmu}
\end{figure}

\noindent \textbf{Multiple displacements}. Multiple applications of the displacement operator break the symmetry of our protocol, by applying a phase to the bottom fringe---the lowest in phase quadrature---see e.g.\ the middle panels in the top row in Fig.~3. To see this, note that the initial state at the origin of phase space is first displaced in the negative $X$ and then in the negative $P$ direction in Fig.~3. Consequently, the top and side fringes originate from a single application of the displacement operator, while the bottom fringes are the result of two applications, which acquire an additional phase due to the relation $\mathcal{D}(\alpha)\mathcal{D}(\beta) {=} e^{\frac{1}{2}(\alpha\beta^* - \alpha^*\beta)}\mathcal{D}(\alpha {+} \beta)$. This phase change becomes more obvious as the fringe wavelength increases with decreasing separation between the coherent states. Figure~\ref{fig:lowtemplowmu} shows how this effect results in a rotation for higher order hypercube states in the regime of vanishing coupling strength.\\

\begin{figure}[!t]
\begin{center}                
\includegraphics[width=\columnwidth]{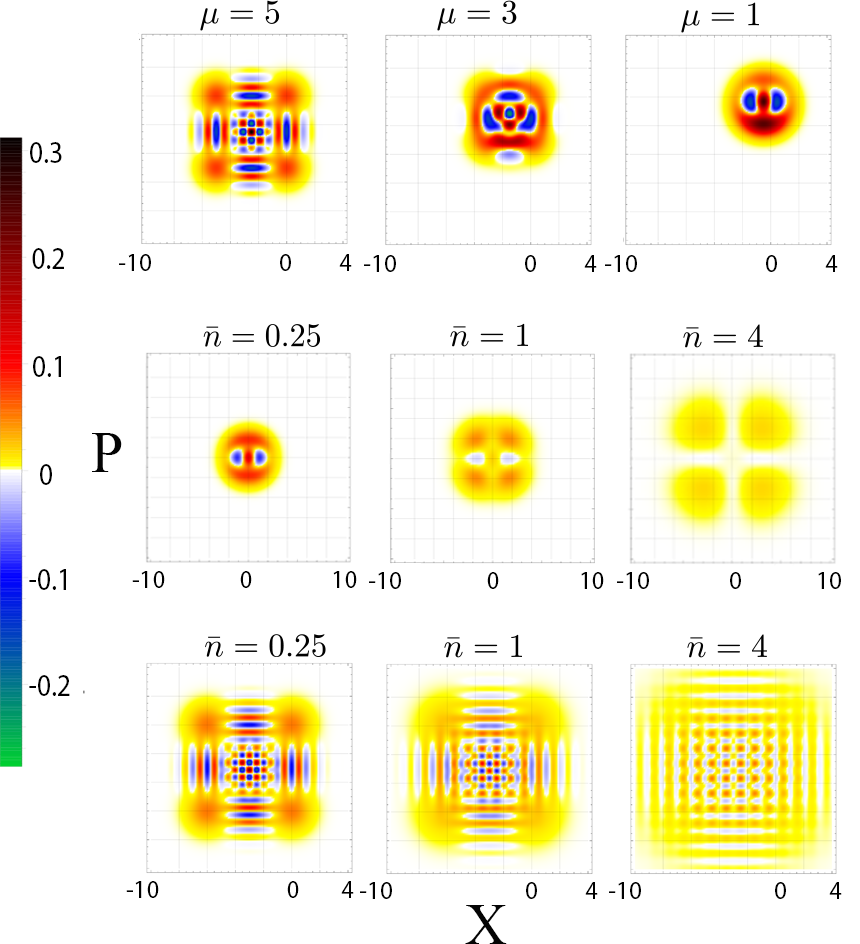}
\end{center}
\vspace{-5mm}
\caption{Wigner function plots for a $n=2$ hypercube state. \textbf{Top row}. Cold temperature: large $\rightarrow$ small coupling ($\bar{n}{=}0$; $\mu{=}5 \rightarrow \mu{=} 1$). \textbf{Middle row} small coupling: cold $\rightarrow$ hot temperature ($\mu{=} 8 {\times} 10^{-9}$; $\bar{n}{=}0.25 \rightarrow \bar{n}{=} 4$; the colours in the last panel are scaled by $10^{8}$ in accordance with the axis scaling). \textbf{Bottom row} Large coupling: cold $\rightarrow$ warm temperature ($\mu{=} 6$; $\bar{n}{=}0.25 \rightarrow \bar{n}{=} 4$). \vspace{-5mm}}
\label{fig:transitons}
\end{figure}

\noindent \textbf{Alternative quantitative indicators for sensing}. We outline the use of the Wigner negativity to quantify the sensitivity of hypercube states to small displacements.

\begin{figure}[!t]
\begin{center}
\includegraphics[width=1.0\columnwidth]{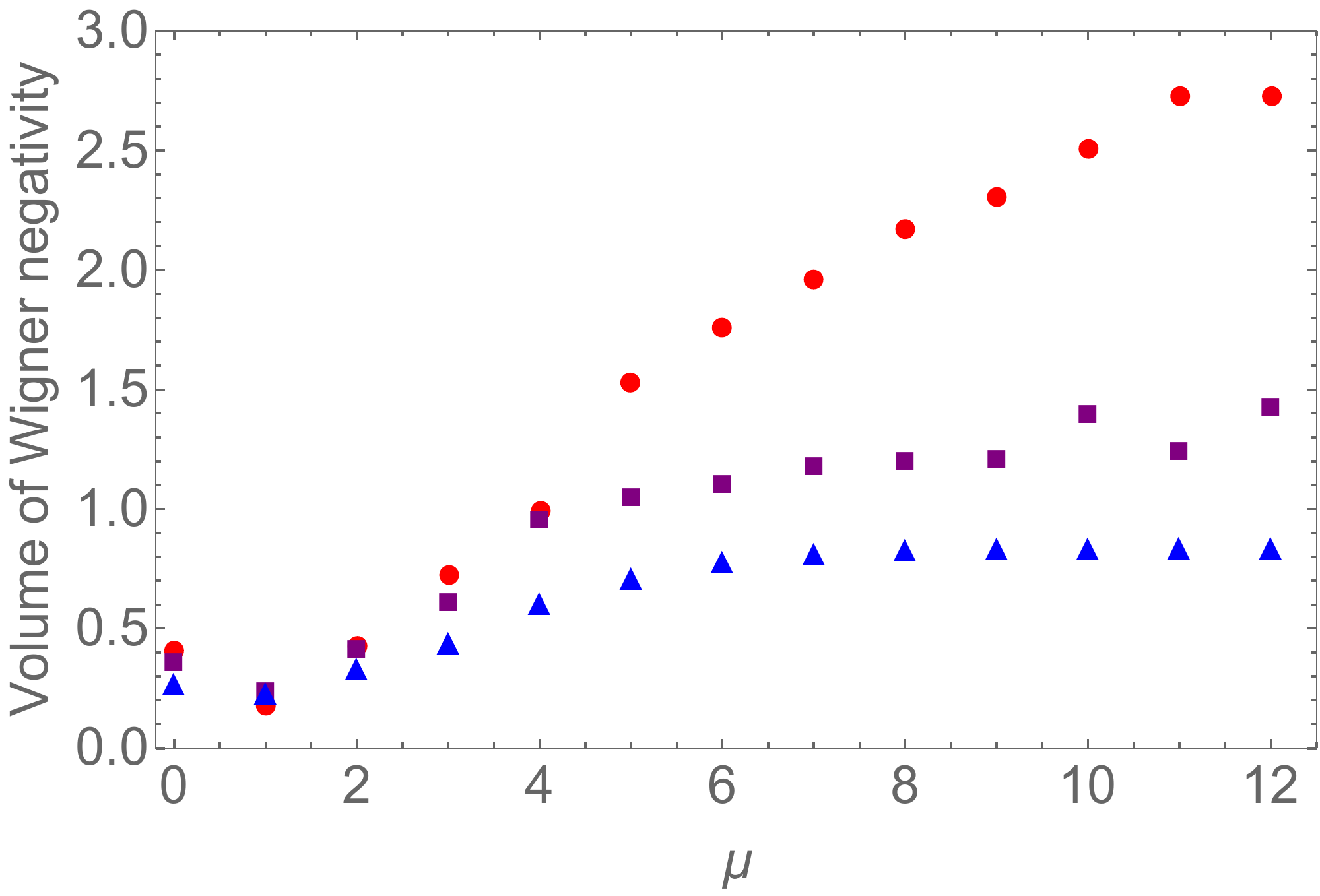}
\includegraphics[width=1.0\columnwidth]{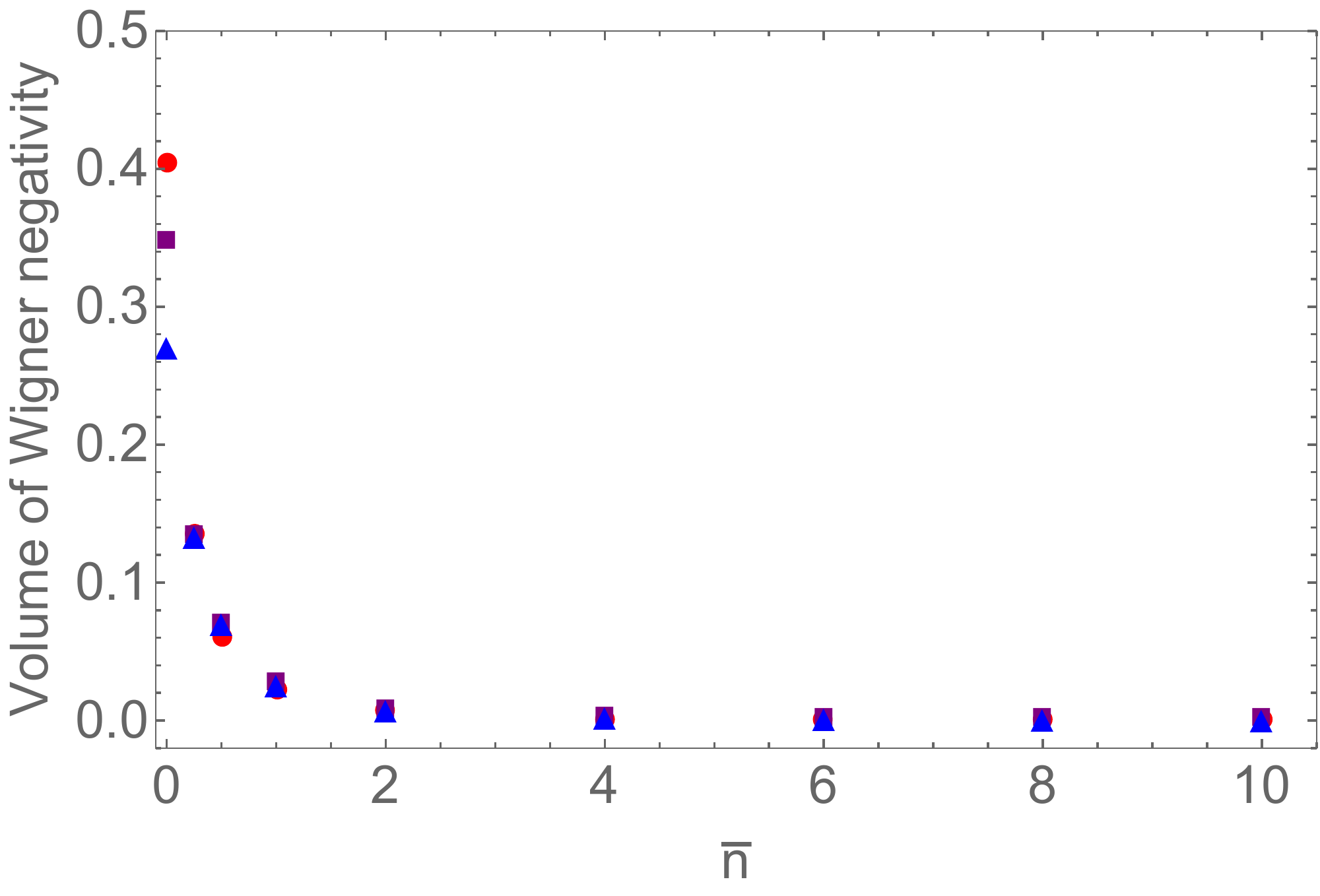}
\includegraphics[width=1.0\columnwidth]{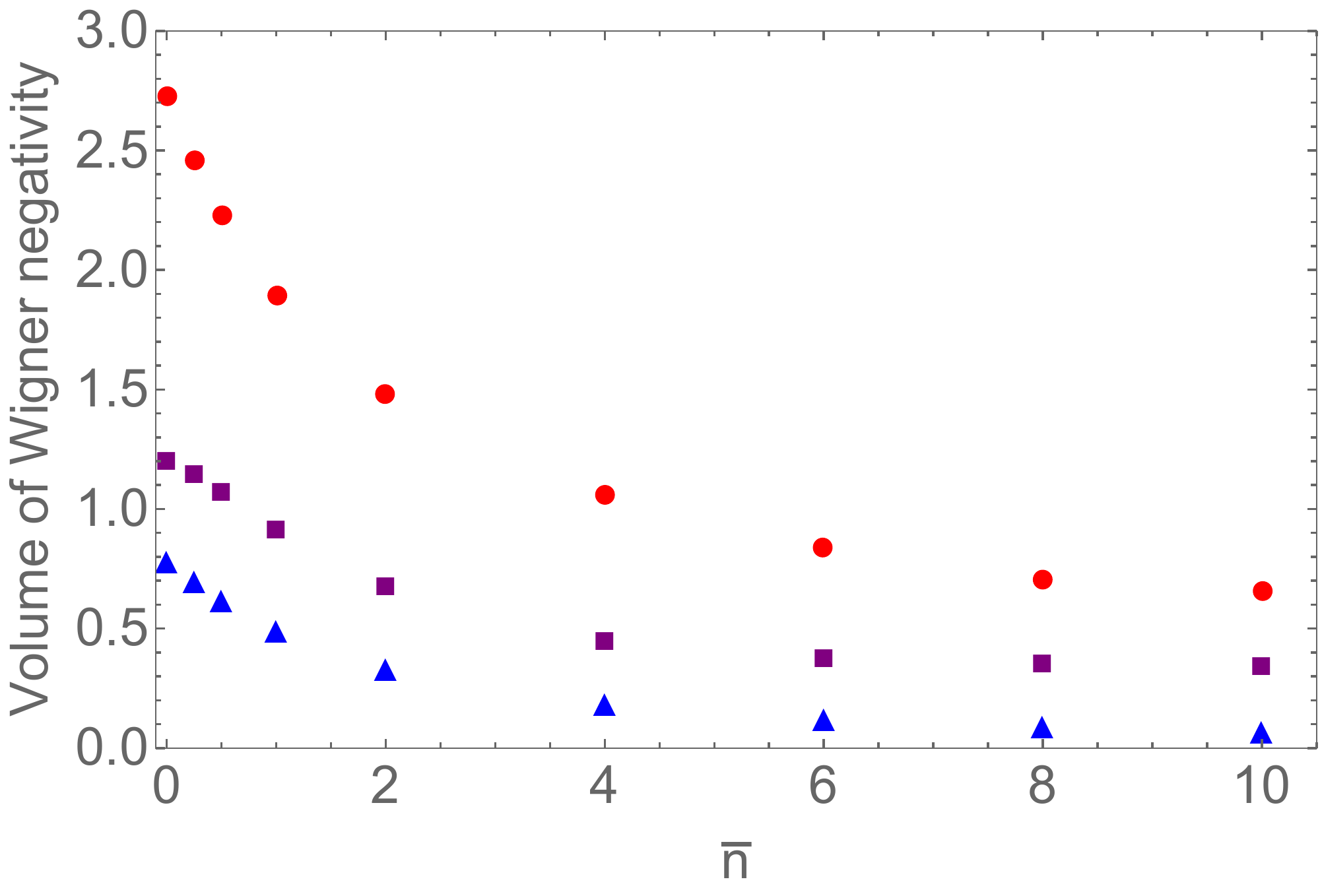}
\end{center}
\vspace{-5mm}
\caption{Volume of Wigner negativity for second- (blue triangles), third- (purple squares) and fourth-order (red circles) hypercube states. \textbf{Top Row}. Cold temperature: large $\rightarrow$ small coupling ($\bar{n}{=}0$; $\mu{=}6 \rightarrow \mu{=} 8 {\times} 10^{-9}$). \textbf{Middle Row} Small coupling: cold $\rightarrow$ hot temperature ($\mu{=} 8 {\times} 10^{-9}$; $\bar{n}{=}0 \rightarrow \bar{n}{=} 10^{15}$; the colours in the last panel are scaled by $10^{8}$ in accordance with the axis scaling). \textbf{Bottom Row} Large coupling: cold $\rightarrow$ warm temperature ($\mu{=} 6$; $\bar{n}{=}0 \rightarrow \bar{n}{=} 6$). \vspace{-5mm}}
\label{fig:Negativity}
\end{figure}

The volume of Wigner negativity is a commonly used measure of the quantumness of a state~\cite{Kenfack2004} with an operational significance as a necessary condition for obtaining quantum computational power~\cite{Veitch2013}. In Fig.~\ref{fig:Negativity} we calculate the volume of Wigner negativity for the second-, third- and fourth- order hypercube states for the conditions of each row from Fig.~3. We see that for large coupling, the Wigner negativity increases significantly with the order of the state (top row) and persists for large $\bar{n}{=}10$ (bottom row). Even for small coupling, $\mu {=} 8 {\times} 10^{-9}$, the negativity is evident for temperatures $\bar{n}{<}2$ (middle row). Fig.~3 highlights that even in this cold small coupling limit Wigner negativity is a prominent feature of hypercube states.\\

\begin{figure}[t]
\begin{center}
\includegraphics[width=\columnwidth]{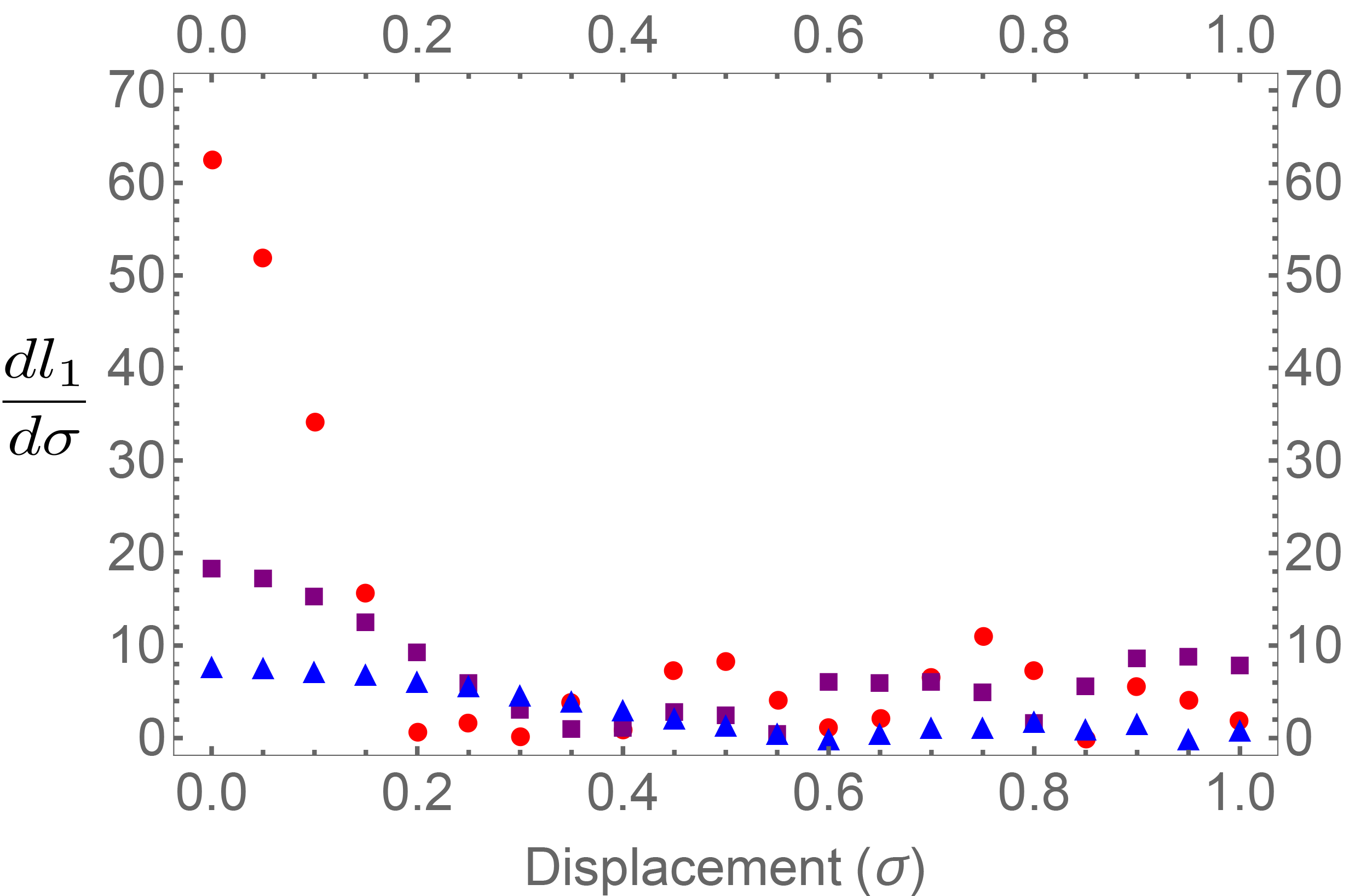}
\end{center}
\vspace{-2em}
\caption{The $\ell_1/d\sigma$ between displaced and undisplaced hypercube states in the momentum quadrature, in terms of the mechanical ground-state width $\sigma$. Hypercube states shown are: compass ($n{=}2$, blue triangles); cube ($n{=}3$, magenta squares); and tesseract ($n{=}4$, red circles).}
\label{fig:L1norms}
\end{figure}

In fig.~\ref{fig:L1norms} we show the $\ell_1/d\sigma$ between displaced and undisplaced hypercube states in the momentum quadrature vs displacement. Hypercube states become increasingly more sensitive with order.

\noindent \textbf{Experimental details \& data analysis}. The mechanical resonator used in our experiment is a high stress $1.7 {\times} 1.7$ mm $Si_{3}N_{4}$ membrane encased in a $10 {\times} 10$ mm silicon frame. The membrane has a thickness of $50\pm2.5$ mm and a reflectivity of $23\pm 0.5\%$, at a wavelength of 795 nm. At the same wavelength the frame has a reflectivity of $20.5 {\pm} 0.2\%$. The resonance frequency of the membrane is $105.64$kHz, with the FWHM of the noise-power spectrum being $3.1 {\pm} 0.05$ kHz. The membrane was driven using a Steminac SM412 ring-piezo which has a nominal resonance frequency of 1.7 MHz and an operating capacitance of $1.8$ nF. 

Photon detection signals from the avalanche photodiode are split, delayed and fed into a time-tagging coincidence logic. For example, to detect creation of the $n {=} 2$ hypercube state, we split the signal into two logical paths, one of which takes time $T/4$ longer for the signal to traverse. Two photons separated by $T/4$ will then create a coincidence event. This coincidence event triggers the recording of a time trace of the mechanical position via homodyne measurement, which can be used to reconstruct the phase space position of the mechanical resonator.

Detection of the third- and fourth- order hypercube states requires conditioning on three and four photons, respectively, which requires splitting the signal into three and four logical paths respectively. The specific timing between photons for the second, third, and fourth order hypercube states was ($2.36  \mu$s), ($1.57  \mu$s, $1.57\mu$s), and, ($1.24 \mu$s, $1.12 \mu$s, $1.18 \mu$s), respectively. (The timing was not equal for the fourth-order hypercube state due to cabling limitations).

Upon finding the requisite coincidence event the logic triggers an oscilloscope which records a trace of the homodyne signal measuring the membranes position. This trace consists of 2500 points sampled at a rate of 100MS/s resulting in a 25$\mu s$ window around the trigger event. The $X$-, $P$- and $\phi$- values for each trace were obtained from a fit of the mechanical response function and the phase space distribution was reconstructed from all fits. The function used for the fitting was,
\begin{align}
&1-d \Big| \cos{ \Big[\omega_{M}t {+} \arctan(X/P) {-} \frac{\pi}{4} \Big) \Big| } \\ \nonumber
& {\times} A \cos \Big[ \big( X \cos (\omega_{M}t) {+} P\sin (\omega_{M}t) {+} \phi \big) {+} c\Big]
\end{align}
where A is the amplitude of the homodyne signal, $\omega_{M}$ is the resonance frequency of the mechanical resonator, $\phi$ is the static phase shift along the arm of the interferometer incident on the membrane frame, $c$ is the DC-component in the signal due to the loss along each arm of the interferometer being asymmetric and $d$ describes the strength of the membrane amplitude modulation that occurs at larger amplitudes. The phase space readout used a balanced photodetector with a gain of $10^5$ V/A and a bandwidth of 4 MHz. The DC component $c$ was compensated to zero by adding an adjustable loss element in front of one detector. Since $\omega_{M}$, $A$, $c$, $d$ were measured independently and constant during measurement, and the other variables, namely, $X$, $P$ and $\phi$, all could be used to find unique, repeatable and accurate fits to the signal.\\

\noindent \textbf{Scaling of the }$\mathbf{\ell_{1}}$\textbf{-sensitivity.}
An analytical calculation of the scaling of the $\ell_1$-sensitivity is very challenging and requires independent rigorous mathematical investigations. In short, the Wigner function of an arbitrary hypercube state generically contains generalized Hermite polynomials, which give rise to the oscillatory features of the Wigner functions in phase-space. In order to obtain the scaling properties of $\ell_1$-sensitivity, we would need to know the behaviour of the zeros of such polynomials upon displacements. In contrast, a numerical calculation of the $\ell_1$-sensitivity is straightforward. Consequently, the $\ell_{1}$-sensitivity results in the main text have been obtained numerically.
By fitting numerically computed points, (such as those of Fig.~3 in the main text), we can then approximate the scaling properties of the $\ell_{1}$-sensitivity or $\mathrm{d}\ell_{1}/\mathrm{d}\sigma$ to
\begin{equation}
\begin{split}
        \mathrm{d}\ell_{1}/\mathrm{d}\sigma \approx & (1.164n+0.046)e^{(-0.526n+0.248)\bar{n}} \\
        &+e^{-0.795\mu}(2.076-1.377n)\mu^2\\
        &+(-0.485+0.698n)\mu\\
        &+e^{-0.807\mu}-0.063n-0.776 \, .
\end{split}
\label{eq:supp:scaling}
\end{equation}
In the experimentally relevant regime of $2\le n\le 4$, $\bar{n} \lesssim 10$ and $\mu\lesssim 8$ \ref{eq:supp:scaling} provides a good approximation of $\mathrm{d}\ell_{1}/\mathrm{d}\sigma$, as quantified by a $\chi^{2}$ value of $0.29$. 

A few interesting features are an initial approximately quadratic increase in sensitivity with $\mu$. For larger $\mu$, where the coherent states no longer overlap, the increase becomes linear with $\mu$. For fixed coupling strength, we also find an exponential decay in the $\ell_1$-sensitivity with increasing temperature, which is valid for a wide range of parameters.

\end{document}